\documentclass[aps, prd, onecolumn, notitlepage, nofootinbib, 12pt]{revtex4-1}
\usepackage{graphicx}
\usepackage{float}
\usepackage{amsmath}
\usepackage{color}
\usepackage{tensor}
\usepackage{epstopdf}
\usepackage{xcolor}
\usepackage{hyperref}
\hypersetup{
    colorlinks,
    linkcolor={blue!50!black},
    citecolor={red!50!black},
    urlcolor={blue!80!black}
}

\usepackage{multibib}

\newcommand{\be}{\begin{equation}}
\newcommand{\ee}{\end{equation}}
\newcommand{\ba}{\begin{eqnarray}}
\newcommand{\ea}{\end{eqnarray}}

\newcommand{\beq}{\begin{equation}}
\newcommand{\eeq}{\end{equation}}
\newcommand{\beqa}{\begin{eqnarray}}
\newcommand{\eeqa}{\end{eqnarray}}
\newcommand{\nn}{\nonumber}

\newcommand{\ca}{\mathcal{A}}
\newcommand{\cb}{\mathcal{B}}

\begin{document}

\title{Universality for Black Hole Heat Engines Near Critical Points}

\author{Maria C. DiMarco}
\email{mdimarco@uwaterloo.ca}
\affiliation{Department of Physics and Astronomy, University of Waterloo, Waterloo, Ontario, Canada, N2L 3G1}

\author{Sierra L. Jess}
\email{sjess@uwaterloo.ca}
\affiliation{Department of Physics and Astronomy, University of Waterloo, Waterloo, Ontario, Canada, N2L 3G1}

\author{Robie A. Hennigar}
\email{robie.hennigar@icc.ub.edu}
\affiliation{Departament de F{\'\i}sica Qu\`antica i Astrof\'{\i}sica, Institut de
Ci\`encies del Cosmos,
 Universitat de
Barcelona,  Mart\'{\i} i Franqu\`es 1, E-08028 Barcelona, Spain}

\author{Robert B. Mann}
\email{rbmann@uwaterloo.ca}
\affiliation{Department of Physics and Astronomy, University of Waterloo, Waterloo, Ontario, Canada, N2L 3G1}
\affiliation{Perimeter Institute, 31 Caroline St. N., Waterloo,
Ontario, N2L 2Y5, Canada}

%


\begin{abstract}
Johnson has shown~\cite{Johnson:2017hxu} that in the vicinity of a critical point the efficiency of a black hole heat engine can approach the Carnot efficiency while maintaining finite power. We characterize and extend this result in several ways, and we show how the rate of approach to the Carnot efficiency is governed by the critical exponents. We apply these results to several classes of black holes to illustrate their validity. Odd-order Lovelock black holes are known to have isolated critical points for which the critical exponents differ from the mean field theory values, providing a non-trivial test of the results. In this case, our results indicate the impossibility of even-order Lovelock black holes with isolated critical points in this class: their existence would constitute a violation of the second law of thermodynamics.
\end{abstract}

\maketitle

\section{Introduction}

Since its inception nearly half a century ago, black hole thermodynamics has provided an arena where quantum effects in gravity can be fruitfully explored. The last decade has witnessed a number of new developments in this field concerning the thermodynamics of gravity with a cosmological constant~\cite{Kastor:2009wy, Cvetic:2010jb}. In this context, the notion of thermodynamic volume naturally arises through regularization of mass in spacetimes with anti-de Sitter asymptotics. The conjugate quantity to the thermodynamic volume is the cosmological constant, interpreted as a pressure. The study of the implications of the thermodynamic volume and the pressure constitute what has come to be known as \textit{black hole chemistry}~\cite{Kubiznak:2016qmn}.

Early results in black hole chemistry included the conjecture of the reverse isoperimetric inequality~\cite{Cvetic:2010jb, Hennigar:2014cfa}, which proposes an upper limit on black hole entropy in terms of thermodynamic volume. Later developments extended these ideas beyond the context of black holes, including smooth geometries and cosmological horizons~\cite{Dolan:2013ft, Mbarek:2016mep,Appels:2016uha, Bordo:2019tyh, Andrews:2019hvq}. Perhaps the most prolific body of work has concerned phase transitions in the pressure-volume phase space. While the study of black hole phase transitions was initiated much earlier~\cite{Hawking:1982dh,Chamblin:1999tk,Caldarelli:1999xj}, in black hole chemistry they are particularly rich. The first example was the rather precise correspondence between the properties of the van der Waals fluid and the thermodynamics of the charged AdS black hole~\cite{Kubiznak:2012wp}. Subsequent studies uncovered a number of similarities between the thermodynamics of black holes and ordinary systems including examples such as triple points, polymer-type phase transitions, superfluid-like phase transitions, and multicriticality, among others~\cite{Altamirano:2013ane, Altamirano:2013uqa, Wei:2014hba, Frassino:2014pha, Dolan:2014vba, Hennigar:2016xwd, Tavakoli:2022kmo,Wu:2022bdk}. We refer to~\cite{Kubiznak:2016qmn} and references therein for a review of results in this area.

The first approach to understand black hole chemistry in the framework of the AdS/CFT correspondence was the notion of a holographic heat engine introduced by Johnson~\cite{Johnson:2014yja}.\footnote{See~\cite{Kastor:2014dra, Karch:2015rpa, Caceres:2016xjz, Couch:2016exn,Johnson:2019wcq, Rosso:2020zkk, AlBalushi:2020rqe, Cong:2021fnf} for other relevant work interpreting black hole chemistry in AdS/CFT.} The idea is to take the black hole through an engine cycle defined in the pressure-volume phase space. The CFT perspective is a cycle defined on the space of dual theories. A number of results have been obtained for holographic heat engines, including exact results for the efficiency~\cite{Johnson:2016pfa, Hennigar:2017apu}, the effects of higher-curvature corrections in the bulk~\cite{Johnson:2015ekr}, and explicit connections between the efficiency of the engine and the ratio of central charges in the dual description~\cite{Johnson:2018amj}. Among the many other results in this subject~\cite{Caceres:2015vsa,Chakraborty:2016ssb, Johnson:2017ood, Chakraborty:2017weq,Zhang:2018vqs, Rosso:2018acz, Yerra:2018mni, Zhang:2018hms, Ghaffarnejad:2018gtj, Johnson:2019olt, Johnson:2019ayc, Ahmed:2019yci, Yerra:2020bfx, ElMoumni:2021zbp, Ferketic:2022ngy}, here we will be interested in a result of Johnson concerning holographic heat engines in the vicinity of a critical point~\cite{Johnson:2017hxu}.

Johnson considered four-dimensional, charged AdS black holes, which exhibit a critical point in the pressure-volume phase space characterized by mean field theory critical exponents. It was demonstrated that a heat engine placed in the vicinity of this critical point has an efficiency that approaches the Carnot efficiency, while maintaining finite power. The critical point itself is important for this process to work, as it dictates the scaling of thermodynamic quantities in its vicinity.  Johnson's result for the ratio of efficiency $\eta$ of these engines to the Carnot efficiency $\eta_C$ took the simple form
\be 
\frac{\eta}{\eta_C} = 1 - \frac{8x}{57} + \mathcal{O}(x^2) \, ,
\ee
where $x$ is a small parameter, taken in this case to be the ratio of the black hole charge to a fixed length scale determined from the critical parameters.

Our aim here is to explain the origin of various aspects of this formula at a deeper level. We show a number of results: 
\begin{enumerate}
\item That the limit $\eta/\eta_C \to 1$ is quite general, applying at least to any black hole exhibiting critical behaviour that also has vanishing heat capacity at constant volume. 
\item We relate the numerical coefficient of the sub-leading term to geometric information of the shape of the engine cycle and the van der Waals ratio of the critical thermodynamic parameters $T_c/(v_c P_c)$, which is universal for most black holes. 
\item Most interestingly, we relate the rate at which the Carnot efficiency is approached, i.e. the power of $x$ that controls the subleading term, directly to the universality class of the critical point. In other words, it is the critical exponents that govern the rate at which the efficiency approaches the Carnot value.
\end{enumerate} 
After establishing these results in the first part of the paper, we then illustrate them for a wide class of black holes. First, we consider charged AdS black holes in all dimensions, and then we consider Lovelock black holes, which include a special class of solutions for which the critical exponents differ from the mean field theory values.

\section{General considerations}\label{sec:general_considerations}

When analyzing the critical behaviour of a black hole, it is common practice to expand the equation of state $P = P(T, V)$ in the vicinity of the critical point. Such an expansion typically takes the form,
\be 
\rho = \sum_{i, j} a_{i,j} \tau^i \omega^j \, ,
\ee
where $a_{i,j}$ will be theory-dependent constants, while the dimensionless parameters $\rho, \tau$ and $\omega$ are related to the corresponding dimensionful thermodynamic parameters via
\be 
P = P_c (\rho + 1) \, , \quad V = V_c (\omega + 1) \,  , \quad T = T_c(\tau + 1) \ .
\ee

For the purposes of calculating the critical exponents --- and further for our purposes --- the above expansion can be simplified and cast in a more useful form. In the computation of the critical exponents, only the leading-order behaviour is important. One can go through the computation of the critical exponents from a general near-critical expansion of the equation of state, as for example in~\cite{Kubiznak:2012wp}, and identify the relevant pieces of the near critical expansion in terms of the critical exponents. Doing so, one finds the following  \textit{near critical expansion} 
\be\label{eqn:near_crit} 
\rho = A\tau + B_0 \tau \omega^{\delta - \frac{1}{\beta}} + C_0 \omega^\delta + \cdots  .
\ee 
The exponents $\beta$ and $\delta$ appearing in the expansion are two of the critical exponents governing how the system behaves as the critical point is approached --- see \cite{Kubiznak:2012wp} for a detailed discussion of the physical meaning of the exponents.  Using the Widom relation allows one to deduce the critical exponent,
\be 
\gamma = \beta(\delta - 1) \, ,
\ee
while the remaining critical exponent $\alpha$ cannot be determined from the equation of state alone, and requires a study of the specific heat at constant volume --- it will vanish in the case $C_V = 0$.  

The first term in the expansion \eqref{eqn:near_crit}  is related to the presence of  ideal gas behaviour in the equation of state. The coefficient $A$ is given by the reciprocal van der Waals ratio, 
\be 
A = \frac{T_c}{v_c P_c} \, .
\ee  
It is possible for $A$ to vanish, but by definition $B_0$ and $C_0$ cannot. In this work, we will see that this expansion suffices to capture the leading order corrections to the efficiency.

Our goal here is to consider how much of Johnson's result for the efficiency of a near critical heat engine is determined from the constants appearing in this relation and the critical exponents.  We will henceforth assume that $C_V = 0$, and more specifically that the entropy is a function of the thermodynamic volume $S = S(V)$. This simplifies the analysis, and is true for most spherically symmetric black holes in Einstein gravity, its higher-curvature extensions, and in situations with matter. Rotating black holes are an example where this assumption would not be true.

\begin{figure}
\centering
\includegraphics[width=0.45\textwidth]{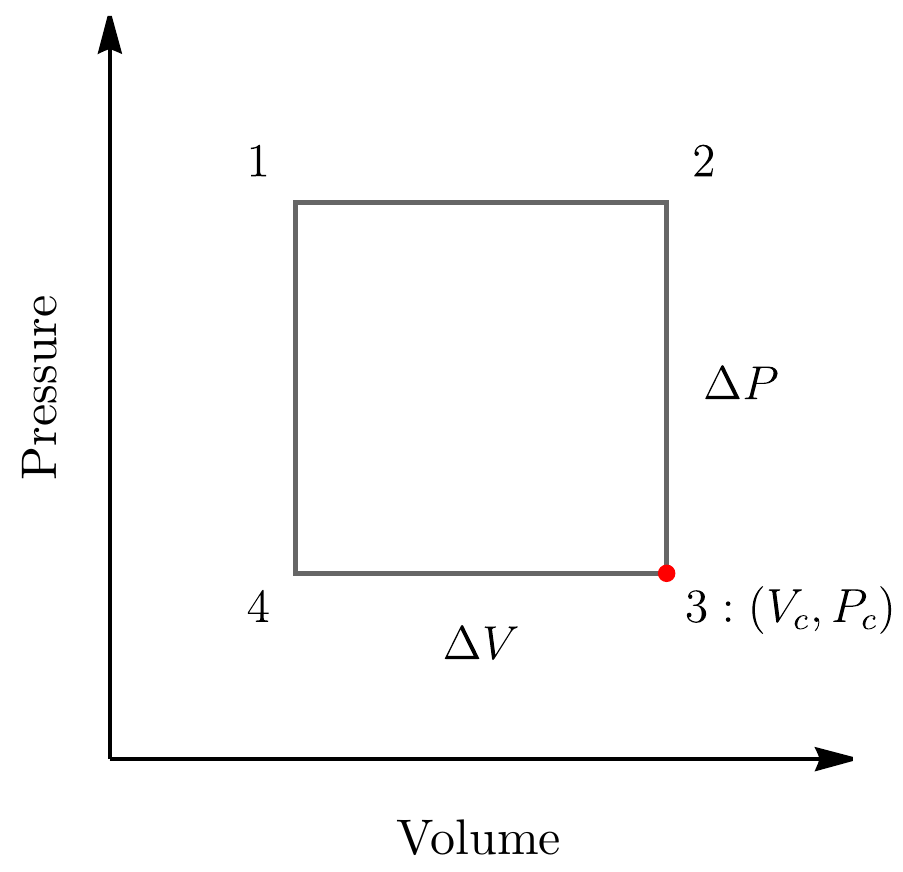}
\caption{A sketch of the engine cycle used in this work. The red dot (lower right corner of the cycle) corresponds to the critical point. }
\label{cycle_diagram}
\end{figure}

For ease of comparison, our setup mirrors that of Johnson. We take a rectangular engine cycle, arranged such that the critical point coincides with the bottom right corner of the rectangle --- see Figure~\ref{cycle_diagram} for a sketch of the cycle with the relevant data labelled.  We take the height of the rectangle to be $\Delta P$ and the width $\Delta V$. Then the efficiency of the cycle is given by~\cite{Johnson:2015ekr}
\be\label{eqn:efficiencyformula} 
\eta = \frac{W}{Q_H} = \frac{\Delta P \Delta V}{M_2 - M_1}
\ee
where we have made use of the fact that $C_V = 0$.  The difference in mass appearing in the denominator can be written in the form of a Taylor series,
\begin{align}
M_2 - M_1 &= M(P_c + \Delta P, V_c) - M(P_c + \Delta P, V_c - \Delta V)
\nn\\
&= \frac{\partial M}{\partial V}\Delta V + \frac{\partial^2 M}{\partial P\partial V} \Delta P \Delta V - \frac{1}{2} \frac{\partial^2 M}{\partial V^2} \Delta V^2 + \cdots,
\end{align} 
where each term is evaluated at the critical point.  We immediately note that derivatives involving only the pressure will be absent from the series, since the pressure is the same at each point.  Furthermore, there are two thermodynamic facts that simplify the above expression.  First, we have, by definition,\footnote{This equality makes use of the assumption that $S = S(V)$.}
\be 
\left(\frac{\partial M}{\partial P} \right)_V = V \, .
\ee
This leads to significant simplification, since it implies that
\be 
\frac{\partial^{m+n} M}{\partial V^n \partial P^m} = 0 \quad \text{for  $m \ge 1$ and $n \ge 2$.}
\ee
This means that the only mixed derivative that contributes to the expansion is
\be 
\frac{\partial^2 M}{\partial P\partial V} = 1\, .
\ee

The second useful thermodynamic identity is
\be\label{ident} 
\left(\frac{\partial M}{\partial V} \right)_P = \left[\left(\frac{\partial}{\partial P }\right)_S \log T \right]^{-1} \, .
\ee
Since we are working under the assumption that $C_V = 0$, this means that the constant entropy derivative can be replaced with a constant volume derivative.  That is,
\be 
\left(\frac{\partial M}{\partial V} \right)_P = \left[\left(\frac{\partial}{\partial P }\right)_V \log T \right]^{-1} \quad \text{for $C_V = 0$.}
\ee
The remarkable feature of this identity is that it allows us to compute the volume derivatives of $M$ using the near critical expansion, which relates the pressure and temperature near criticality.  For computing purposes, we note that by the chain rule,
\be  
\frac{\partial }{\partial V} = \frac{1}{V_c} \frac{\partial }{\partial \omega} \, , \quad  
\frac{\partial }{\partial T} = \frac{1}{T_c} \frac{\partial }{\partial \tau} \, , \quad
\frac{\partial }{\partial P} = \frac{1}{P_c} \frac{\partial }{\partial \rho} \,.
\ee

Using the near critical expansion~\eqref{eqn:near_crit}, we find that
\be
\left(\frac{\partial M}{\partial V} \right)_P = P_c \left(A + \rho + B_0 \omega^{\delta - \frac{1}{\beta}} - C_0 \omega^\delta \right) \, .
\ee

Further, the first higher order derivative that is non-vanishing when evaluated at the critical point can be found to be
\be 
\left(\frac{\partial^{n+1} M}{\partial V^{n+1}} \right)_P = \begin{cases} 
n!  \dfrac{ B_0 P_c}{V_c^n} \quad &\text{if } n = \delta - \frac{1}{\beta}
\\ \\
0 \quad &\text{otherwise} \, .
\end{cases}
\ee
Thus, including only the lowest order contributions in the Taylor series expansion, the efficiency of the cycle is given by
\begin{align} 
\eta &= \frac{1}{P_c} \frac{\Delta P \Delta V}{A \Delta V + \frac{\Delta P \Delta V}{P_c} +  (-1)^{n_{\star}} \frac{V_c B_0}{n_\star + 1} \left(\frac{\Delta V}{V_c}\right)^{n_{\star}+1} }
\nn\\
&= \frac{y}{A  + y  +  (-1)^{n_{\star}} \frac{B_0}{n_\star + 1}  x^{n_{\star}}}
\end{align}
where  
\be\label{eqn:x_y_def}
\Delta P \equiv y P_c \, , \quad \Delta V \equiv x V_c  
\ee
and the integer $n_\star > 0$ satisfies $n_\star = \delta - \frac{1}{\beta}$.

Note that Johnson's large charge limit is equivalent to the limit $x \to 0$ in this notation.  In this limit, finite work can be maintained by making use of the scaling of the thermodynamic quantities at the critical point. We could, for example, identify $x = LM_c^{-1/(d-3)}$ (where $M_c$ is the mass at criticality and $L$ is a fixed constant) and take the limit $M_c \to \infty$ holding $L$ fixed.  Since at criticality $P_c \propto M_c^{-2/(d-3)}$ and $V_c \propto M_c^{3/(d-3)}$, the work will behave as $\Delta P \Delta V = y P_c V_c L$, which is finite in this limit. 

In this case, we find in the $x \to 0$ limit, the efficiency of the cycle is given by
\be\label{eqn:normal_eff} 
\eta = \frac{y}{A + y} - \frac{y}{(A+y)^2} \frac{B_0}{\delta - \frac{1}{\beta} + 1} (-x)^{\delta - \frac{1}{\beta}} + {\cal O}(x^{2\delta - \frac{2}{\beta}}) \, .
\ee
This makes it possible to see precisely how the properties of the black hole and the universality class of the critical point enter into the efficiency of the engine.  In particular, we see that the efficiency limits to $y/(A+y)$, and the fall off rate of the next to leading order term is governed by the critical exponents as $\delta - \frac{1}{\beta}$.  It is interesting that the universality class completely determines how rapidly the efficiency approaches its limiting value.

We can use the same techniques as above to write down the efficiency of the Carnot cycle.  For this, we need to know the temperatures at the top right and bottom left of the cycle.  To get these temperatures, we can again use a Taylor series of the near critical equation of state.  Doing so yields
\begin{align}
T_C &= T(P_c, V_c - \Delta V) 
\nn\\
&= T_c - \left(\frac{\partial T}{\partial V} \right)_P \Delta V + \frac{1}{2} \left(\frac{\partial^2 T}{\partial V^2} \right)_P (\Delta V)^2 + \cdots \, , 
\\
T_H &= T(P_c + \Delta P, V_c)
\nn\\
&= T_c + \left(\frac{\partial T}{\partial P} \right)_V \Delta P + \frac{1}{2} \left(\frac{\partial^2 T}{\partial P^2} \right)_V (\Delta P)^2 + \cdots \, ,
\end{align}
where all derivatives are evaluated at the critical point. Using the near critical equation of state we can evaluate the derivatives.  A simplification occurs because this equation is linear in the pressure: all pressure derivatives $\partial^n T/\partial P^n$ with $n > 1$ vanish.

  This allows us to compute $T_H$ exactly.  Here in our explicit calculations for $T_C$, we will concern ourselves only with the first non-trivial corrections.  This correction depends crucially on whether or not $A$ is zero.  The result is
\be 
\frac{1}{n!} \left(\frac{\partial^n T}{\partial V^n} \right)_{P} = \begin{cases}
-\dfrac{C_0 T_c}{A  V_c^\delta} \, &\text{if } A \neq 0 \text{ with } n = \delta \,,
\\ \\
- \dfrac{C_0 T_c}{B_0 V_c^{1/\beta}} \, &\text{if } A = 0 \text{ with } n = \frac{1}{\beta}.
\end{cases}\label{eq:dtdv}
\ee

Making use of these results, we find the following results for the maximal and minimal temperature on the cycle (assuming that $A \neq 0$):

\begin{align} 
T_C &= T_c \left(1 - (-1)^{\delta }  \frac{ C_0}{A} \left(\frac{\Delta V}{V_c} \right)^\delta + \cdots\right)\, , \label{eq:t_cold}
\\
T_H  &= T_c \left(1 + \frac{\Delta P}{A P_c} \right) \, .\label{eq:t_hot}
\end{align}
From these expressions, the Carnot efficiency $\eta_C = 1-T_C/T_H$ is easily computed as
\begin{align}\label{eqn:carnot_gen}
\eta_C &= \frac{y}{A + y} + \frac{C_0}{A+y} (-x)^{\delta} + \cdots 
\end{align} 
where $x$ and $y$ are as before [cf. Eq.~\eqref{eqn:x_y_def}]. \

Comparing the Carnot efficiency to the efficiency calculated in Eq.~\eqref{eqn:normal_eff}, we notice that the leading order behaviour in the $x \to 0$ limit is identical.  Further, we can see that the Carnot efficiency approaches this value at a faster rate: here the exponent is $\delta$ rather than $\delta - \frac{1}{\beta}$.  

To compare directly with Johnson's work, we compute the ratio of the two efficiencies.  Including only the first correction, we see that
\be\label{eqn:ratio_gen} 
\frac{\eta}{\eta_C} = 1 - \frac{1}{\delta - \frac{1}{\beta} + 1} \frac{B_0}{A+y} (-x)^{\delta - \frac{1}{\beta}}  + \cdots \, ,
\ee
and so the efficiency limits to the Carnot efficiency in the limit $x \to 0$, while finite work is maintained due to the scaling of the temperature, volume, and pressure at the critical point.

Note that the knowledge flows both ways here: Not only does the near critical equation of state inform us about the efficiency, but the expression for the efficiency also informs us about the near critical equation of state. For example, the constraint that the efficiency should always be less than one tells us that $A \ge 0$ always. Furthermore, if $\delta - \frac{1}{\beta}$ is odd, then $B_0$ should be negative and if $\delta - \frac{1}{\beta}$ is even, then $B_0$ should be positive.  

We see that an analysis of heat engines provides a (much simpler) means to evaluate the critical exponents. A computation of the Carnot efficiency allows for a determination of $\delta$ from the leading-order correction to the limiting efficiency. Once $\delta$ has been determined in this way, $\beta$ can be extracted from the efficiency of the rectangular cycle, as the exponent of the correction in this case is $\delta-1/\beta$. The Widom relation can then be used to obtain $\gamma$, giving the full set of critical exponents.

We will now consider examples. We will compare the result of a direct computation of the efficiency with the result of our results derived above, showing directly its validity. We will consider two main examples. First, we will study the charged AdS black hole in all dimensions. This can be viewed as an extension of Johnson's work. Then we will consider a particular class of Lovelock black holes that exhibit isolated critical points. These are, to the best of our knowledge, the only known black holes that have critical exponents that differ from the mean field theory values. Therefore, an analysis of these solutions provides a non-trivial test of the results that we have derived and the role of the critical exponents in determining the approach to maximal efficiency. In an appendix, we present an additional example for black holes in Einstein-Gauss-Bonnet-Maxwell theory. While they belong to the same universality class as charged AdS black holes, the equation of state is more complicated, making for a messier but useful test.

\section{Example 1: RN-AdS in all dimensions}\label{sec:RN_AdS}

The extended thermodynamics of charged AdS black holes in all dimensions was studied in~\cite{Gunasekaran:2012dq}. Here we will use the relevant results from that paper without derivation. In $d$ spacetime dimensions, the metric for the charged AdS black hole takes the form
\be \label{static_symmetric_metric}
ds^2 = - f dt^2 + \frac{dr^2}{f} + r^2 d \Omega^2_{d-2}
\ee
where 
\be 
f = 1 - \frac{m}{r^{d-3}} + \frac{q^2}{r^{2(d-3)}} + \frac{r^2}{\ell^2} \, .
\ee
The mass can be expressed in terms of $r_+$ by solving the condition $f(r_+) = 0$, yielding
\begin{align} 
M &= \frac{(d-2) \omega_{d-2}}{16 \pi} m
\nn\\
&= \frac{(d-2) \omega_{d-2}}{16 \pi} \left[ (\kappa v)^{d-3} + \frac{16 \pi (\kappa v)^{d-1} P}{(d-1)(d-2)} + \frac{q^2}{(\kappa v)^{d-3}}  \right], 
\end{align}
where $v$ is the specific volume defined as
\be 
v = \frac{1}{\kappa} \left[\frac{(d-1) V}{\omega_{d-2}}\right]^{1/(d-1)} \, ,
\ee
with $V$ --- the thermodynamic volume --- given in terms of $r_+$,
\be 
V = \frac{\Omega_{d-2} r_+^{d-1}}{d-1},
\ee
and $\Omega_{d-2}$ being the volume of the unit sphere in $d-2$ dimensions,
\be\label{eqn:omega(d-2)} 
\Omega_n = \frac{2 \pi^{\frac{n+1}{2}}}{\Gamma\left(\frac{n+1}{2} \right)} \, .
\ee
The pressure is related to $\ell^2$ by
\be
\ell^2 = \frac{(d-1)(d-2)}{16\pi P} \, .
\ee
We have also introduced the factor
\be 
\kappa = \frac{d-2}{4}
\ee
to simplify the expressions.

The equation of state can be obtained by solving the expression  $T = f'(r_+)/4\pi$ for the pressure
in terms of the other thermodynamic quantities
\be 
P = \frac{T}{v} - \frac{(d-3)}{\pi (d-2) v^2} + \frac{q^2(d-3)}{4 \pi v^{2(d-2)} \kappa^{2d - 5}} \, .
\ee
The system has a critical point satisfying
\be 
\frac{\partial P}{\partial v} = \frac{\partial^2 P}{\partial v^2} = 0
\ee
with the critical values,
\begin{align}
v_c &= \frac{1}{\kappa} \left[q^2 (d-2)(2d-5) \right]^{\frac{1}{2(d-3)}} \, ,
\nn\\
T_c &= \frac{(d-3)^2}{\pi \kappa v_c(2d-5)}
\, ,
\nn\\
P_c &= \frac{(d-3)^2}{16 \pi \kappa^2 v_c^2} \,.
\end{align}
While the individual critical values depend on the charge parameter $q$, the van der Waals ratio \be 
\frac{P_c v_c}{T_c} = \frac{2d-5}{4d-8} 
\ee
is universal \cite{Gunasekaran:2012dq}.

We begin by constructing the efficiency of the engine using direct computation.  The cycle is as previously described: rectangular, positioned such that the critical point coincides with the bottom-right corner of the cycle. We take the dimensions of the cycle to be the same as given in Eq.~\eqref{eqn:x_y_def}.

  To calculate the efficiency of the engine then requires determining the mass difference $M_2 - M_1$.  This can of course be computed exactly, but here we will be interested in the small $x$ expansion of this quantity which is,
\begin{align} 
M_2 - M_1 &= \frac{q \Omega_{d-2}}{16 \pi} \frac{(d-3)^2}{d-1} \sqrt{\frac{d-2}{2d-5}} \bigg[ \left( (2d-5) y + 4d - 8\right) x + \frac{2(d-2)}{d-1} x^2 + {\cal O}(x^3) \bigg]
\end{align}
yielding for the efficiency,
\begin{align} 
\eta &= \frac{\Delta P \Delta V}{M_2 - M_1} 
\nn\\
&= \frac{y}{y + (4d - 8)/(2d - 5)} - \frac{2(d-2)}{(2d-5)(d-1)} \frac{y}{\left(y + (4d - 8)/(2d - 5) \right)^2} x + {\cal O}(x^2)
\end{align}

To calculate the Carnot efficiency, we must determine the temperatures $T_C = T_4$ and $T_H = T_2$.  These can be determined exactly via the equation of state.  The expressions are somewhat messy in general dimensions, so we do not write them explicitly.  However, the final result for the Carnot efficiency is
\begin{align}
\eta_C &= 1- \frac{T_C}{T_H} = \frac{y}{y + (4d - 8)/(2d - 5)} + \frac{2d - 4}{3(d-1)^3} \frac{1}{y + (4d - 8)/(2d - 5)} x^3 + {\cal O}(x^4) \, .
\end{align}
We therefore have the ratio
\begin{align} 
\frac{\eta}{\eta_C} &= 1 - \frac{2(d-2)}{(2d-5)(d-1)} \frac{1}{y + (4d - 8)/(2d - 5)} x + {\cal O}(x^2) \, .
\end{align}
Note that, in four dimensions, we recover Johnson's result by choosing $y = 1/2$.  

To compare with the results derived earlier, we must consider the expansion of the equation of state near the critical point.  A simple Taylor expansion reveals
\be 
\rho = \frac{4d - 8}{2d - 5} \tau - \frac{4(d-2)}{(2d-5)(d-1)} \omega \tau - \frac{2}{3} \frac{(d-2)}{(d-1)^3} \omega^3 + \cdots 
\ee
from which we can read off
\begin{align}
A &= \frac{4d - 8}{2d - 5} \, , 
\nn\\
B_0 &= - \frac{4(d-2)}{(2d-5)(d-1)}
\, ,
\nn\\
C_0 &= - \frac{2}{3} \frac{(d-2)}{(d-1)^3} \, ,
\end{align} 
and the critical exponents are
\be \label{crit_exp_standard}
\beta = \frac{1}{2} \, , \quad \gamma = 1 \, , \quad \delta = 3 \, .
\ee

Comparing with expressions~\eqref{eqn:normal_eff}, \eqref{eqn:carnot_gen} and \eqref{eqn:ratio_gen}, it is immediately clear that the results are the same, as expected.  Further, since the equation of state admits an ideal gas limit, we see that the leading order efficiency of both the rectangular and Carnot engines is
\be 
\eta_{\rm LO} = \frac{y}{y + \frac{T_c}{P_c v_c}} \, .
\ee
That is, the leading order behaviour is characterized by one parameter that describes the shape of the cycle and the universal critical ratio. Also, since the critical ratio is a monotonically decreasing function of the spacetime dimension,  the efficiency of any rectangular engine will increase as the spacetime dimension increases.

As described in the previous section, finite work can be achieved in the $x \to 0$ limit only because of the scaling of the thermodynamic quantities with the critical point, since $x$ can be taken to be of the form $L/q$ with $L$ fixed and $q \to \infty$.  Since $P_c \propto q^{-2}$ and $V_c \propto q^3$, this choice guarantees finite work as $q \to \infty$.  

\section{Example 2: Black Holes with Isolated Critical Points}

For our second example we wish to perform a more nontrivial test of the results outlined earlier. To this end, we will study a class of black holes that have critical exponents that differ from the mean field theory values. Such black holes were first reported in~\cite{Dolan:2014vba} and are solutions to the Lovelock class of gravitational theories.

Lovelock gravity~\cite{Lovelock:1971yv} is the most natural higher curvature generalization of Einstein gravity in that it maintains second-order field equations on any metric. At  $k$th-order in the curvature, Lovelock black holes in $d$ dimensions are governed by the action \cite{Cai:2013qga,Frassino:2014pha}
\be
I=\frac{1}{16\pi}
\int d^dx \sqrt{-g}\left(\sum_{k=0}^{k_{\rm max}}\hat{\alpha}_{(k)}\mathcal{L}^{(k)} + \mathcal{L}_m\right) \, ,
\ee
where $\mathcal{L}_m$ denotes any potential matter terms, the $\hat{\alpha}_{(k)}$ are the Lovelock coupling constants, $\mathcal{L}^{(k)}$ are the $2k$-dimensional Euler densities, and $k_{\rm max}$ is the integer part of $(d-1)/2$.\footnote{Explicit expressions for these densities can be found, for example, in~\cite{Frassino:2014pha}.} Thus, the $\hat{\alpha}_0$ term is the cosmological constant term (equivalently, the pressure term) with
\be
\Lambda=-\frac{\hat{\alpha}_0}{2}
\ee and thus
\be
P=\frac{\hat{\alpha}_0}{16\pi}.
\ee
Note that setting $\hat{\alpha}_{1}=1$, $\hat{\alpha}_{(k)}=0,\, k>1$ recovers general relativity with $G = 1$. We will set $\hat{\alpha}_{1} = 1$ for the remainder of the analysis.

We are interested in static and spherically/planar/hyperbolic symmetric black holes in these theories, which in the vacuum case are described by a metric of the form (\ref{static_symmetric_metric}) where $f$ solves the polynomial equation
\begin{align}
&\sum_{k=0}^{k_{max}}\alpha_k\left(\frac{\kappa-f}{r^2}\right)^k=\frac{16 \pi M}{(d-2)\Sigma_{d-2}^{(\kappa)}r^{d-1}},
\end{align}
where $\Sigma_{d-2}^{(\kappa)}$ is equal to the volume of the transverse space. In spherical space, where $\kappa = 1$, it is equal to $\Omega_{d-2}$ as in Eq.~\eqref{eqn:omega(d-2)}.
In the above, the parameter $\kappa$ denotes the topology of the transverse spatial sections. Specifically, $\kappa = \{-1, 0, +1\}$ corresponds to hyperbolic, planar, and spherical topologies, respectively. The $\alpha_k$ are the rescaled Lovelock coupling constants, given by
\be
\alpha_0=\frac{\hat{\alpha}_0}{(d-1)(d-2)},\quad\alpha_1=\hat{\alpha}_1,\quad\alpha_k=\hat{\alpha}_{k}\prod_{n=3}^{2k}(d-n).
\ee

Defining the horizon radius to be $r_+$ (the largest root of $f = 0$), the black hole mass $M$ is given by
\be
M=\frac{\Sigma_{d-2}^{(\kappa)}(d-2)}{16\pi}\sum_{k=0}^{k_{\rm max}}\alpha_k\kappa^kr_+^{d-1-2k},
\ee
The temperature is fixed by demanding regularity of the Euclidean sector and is $T=|f'(r_+)|/4\pi$. The temperature can be rearranged to give the equation of state
\begin{align}\label{Lovelock_EOS}
P=&\frac{d-2}{16\pi}\sum_{k=1}^{k_{\rm max}}\frac{\alpha_k}{r_+^2}\left(\frac{\kappa}{r_+^2}\right)^{k-1}\left(4\pi k r_+ T - \kappa(d-2k-1)\right).
\end{align}

A number of studies have investigated the critical behaviour of Lovelock black holes (see e.g.~\cite{Frassino:2014pha, Cai:2013qga, Hennigar:2016xwd}).   Here our interest lies in understanding how the efficiency of heat engines near critical points in these theories behaves.

In odd-order ($K=2N+1$) Lovelock gravity with the Lovelock couplings fine-tuned such that 
\begin{align}
\alpha_n=\alpha_K \ca^{K-n}\left(
\begin{array}{c}
K\\
n
\end{array}\right)\quad2\le n<K\label{couplings}
\end{align}
with 
\be
\ca\equiv(K\alpha_K)^{\frac{-1}{K-1}} \, ,
\ee
a novel feature emerges for hyperbolic black holes: there is an \textit{isolated critical point} occurring at a thermodynamic singularity \cite{Dolan:2014vba} characterized by non-standard critical exponents
\begin{align}\label{isoCrits}
\beta=1\quad\gamma=K-1\quad\delta=K .
\end{align}
These are the only examples of critical exponents differing from the mean field theory values known for black holes. While isolated critical points have now been observed in other contexts~\cite{Dykaar:2017mba, Hennigar:2016ekz}, all possess these particular non-standard critical exponents.  It is these non-standard critical exponents that will allow us to make a non-trivial check of the results presented in the first section.

The equation of state of such a black hole is
\begin{align}
P=&\frac{(d-1)(d-2) \alpha_K}{16 \pi}\bigg[\cb^{K-1}\left(\frac{2K (2\pi r_+ T+\kappa)}{(d-1) r_+^2}-\cb\right)+\ca^K\bigg]
\end{align}
with 
\be 
\cb\equiv\frac{\kappa}{r_+^2}+\ca \,.
\ee $r_+$ is, as usual, the horizon radius; it can be expressed in terms of the thermodynamic volume $V$ according to
\be\label{volume_radius}
V=\frac{\Sigma_{d-2}^{(\kappa)} r_+^{d-1}}{d-1}.
\ee

It is easy to verify that the equation of state admits a single critical point, with the critical values given by
\begin{align}
r_c=\frac{1}{\sqrt{\ca}}\quad T_c=\frac{1}{2\pi r_c}\quad P_c=\frac{(d-1)(d-2) \alpha_K}{16 \pi}\ca^K \, ,
\end{align}
where $r_c$ denotes the value of the horizon radius at criticality.  Expanding the equation of state near the critical point yields
\be\label{nearcrit_isolated}
\rho=\frac{2^K \,K}{(d-1)^{K}} \tau\omega^{K-1} + \frac{2^K(K-d+1)}{(d-1)^{K+1}}\omega^K+\cdots.
\ee 

Note that this expansion is characterized by $A = 0$. Physically, $A = 0$ is related to the fact that the critical point coincides with the thermodynamic singularity, i.e. $\partial P/\partial T=0$ and thus $\partial T/\partial P\rightarrow\infty$ at the critical point. Consequently  $T(P,V_c)\rightarrow\infty$ for $P\ne P_c$, and   the temperature on the right isochore of the engine cycle we have considered is infinite. 

It is necessary to address whether the analysis of section \ref{sec:general_considerations} is applicable in the $A= 0$ case, as some of the expressions involved terms of the form $1/A$. In doing this, we note that the expression \eqref{eq:t_hot} is indeed valid: $\left(\partial T/\partial P\right)\rightarrow\infty$, and so $T_H\rightarrow \infty$ on the isochore $V=V_c$ as long as $P\ne P_c$. Thus any cycle including this isochore will have a Carnot efficiency of unity,
\be
\eta_C=1.
\ee
The analysis of $\eta$ goes through unmodified as none of the intermediate expressions depend on $1/A$. The expected behaviour of the efficiency is then the same as that given in \eqref{eqn:normal_eff}, and since $\eta_C = 1$, this behaviour captures also the ratio.

Let us apply our result  \eqref{eqn:normal_eff} to the case of the isolated critical point. It is straightforward to carry out the analysis for any value of $K$.  We are led to expect the following behaviour for the efficiency 
\be 
\frac{\eta}{\eta_C} = \eta = 1 - \frac{2^K}{y(d-1)^K} (-x)^{K-1} + \dots \, ,
\ee
where we have extracted the value of $B_0 = 2^K/[K(d-1)^K]$ from~\eqref{nearcrit_isolated} and the critical exponents from~\eqref{isoCrits}.

Let us now compare with a direct computation of the engine efficiency. The mass of these black holes is
\be
M=\frac{(d-2) \Sigma_{d-2}^{(\kappa)} r_+^{d-1}}{16 \pi}\left(\alpha_{K} \cb^K-\alpha_{K} \ca^K+\frac{16\pi P}{(d-1)(d-2)}\right).
\ee
From this, we obtain the the mass difference at the corners of the engine cycle,
\be 
M_2 - M_1 = \frac{(d-2) K^{\frac{d - 2K - 1}{2(K-1)}} \alpha_K^{\frac{d-3}{2(K-1)}} \Sigma_{d-2}^{(\kappa)}}{16 \pi } \left[xy - \left(-\frac{2x}{d-1}\right)^K + \mathcal{O}(x^{K+1})\right]
\ee
The efficiency calculated from Eq.~\eqref{eqn:efficiencyformula} is then
\be\label{isoEff} 
\frac{\eta}{\eta_C} = \eta = 1 - \frac{1}{y} \left(\frac{2}{d-1}\right)^K(-x)^{K-1} + \mathcal{O}(x^K)  
\ee
 in perfect agreement with the result of the general analysis presented just above.

This result also points toward an interesting interpretation of the isolated critical points. Recall these exist only in Lovelock theories of odd order $K$, and this is for good reason. If critical points belonging to the same family (i.e. having the same near-critical equation of state) existed for even values of $K$, then it would be possible to construct a heat engine with efficiency larger than the Carnot efficiency, as evidenced by equation~\eqref{isoEff}. Therefore, if black holes with non-standard critical exponents exist in even-order Lovelock theory, they must differ in some way from those considered here.

\section{Discussion}

We have studied the efficiency of black hole heat engines when the engine cycle is placed in the vicinity of a critical point. This analysis clarifies and extends the work of Johnson, where it was demonstrated that such a configuration allows for an engine efficiency to approach the Carnot limit, while maintaining finite power. We have, for the case of black holes with $C_V = 0$ and rectangular cycles, derived explicit expressions for the efficiency and its ratio to the Carnot efficiency in terms of the cycle geometry, universal thermodynamic data (such as the van der Waals ratio), and the critical exponents. We then illustrated the validity of these results for Einstein-Maxwell black holes and Lovelock black holes. The latter was a particularly interesting case to study, as it allows for non-standard critical exponents which provide a non-trivial test of our results. This also led to an interesting observation concerning why this family of non-standard critical exponents exists only for Lovelock theory of odd order: the same family, extended to even orders, would allow for a violation of the second law of thermodynamics. 

At a more pragmatic level, our results allow for a much easier computation of the critical exponents associated with a second-order phase transition. This can be achieved simply by computing the efficiency of a black hole heat engine in the vicinity of the critical point. Since a simple, exact formula exists for that efficiency~\eqref{eqn:efficiencyformula}, it is a comparatively straight-forward computation.

There are a number of directions in which this work could be extended. In our view, the most natural extension would be to consider black holes with non-vanishing $C_V$. Rotation black holes and STU black holes fall into this class~\cite{Hennigar:2017apu, Johnson:2019vqf}. This extension is not necessarily trivial. In our calculations, we have relied on identities such as Eq.~\eqref{ident}, which would fail to hold in cases where $C_V \neq 0$. One still has a simple expression for the engine efficiency akin to Eq.~\eqref{eqn:efficiencyformula}~\cite{Hennigar:2017apu}, so perhaps a similar analysis can be carried out, utilizing fresh ideas. 

There are a number of other directions worthy of further consideration. It would be interesting to understand how the results depend on the shape of the engine cycle used. To facilitate direct comparison with Johnson's work, all of our results assume that the engine cycle is rectangular. We expect that geometrical properties of the cycle will enter into the coefficients in the $\eta/\eta_C$ expansion. It would be unexpected and surprising to find  that altering the cycle geometry could alter the fact that the critical exponents control the rate of approach. Nonetheless, it would be worthwhile to confirm this carefully. It would of course be of interest to further confirm the validity of the expressions we have derived, testing them against black holes with more complicated equations of state. In this direction, a particularly natural candidate would be the recently discovered examples of \textit{multi}-critical points in black hole physics~\cite{Tavakoli:2022kmo,Wu:2022bdk}. The universality class of these critical points is yet to be determined, so they may provide further examples of critical exponents outside of the mean field theory class. In any case, tuning of the parameters in these models to allow for several of the critical points to coincide should yield non-standard critical exponents in the same sense as~\cite{Dolan:2014vba}. 

Finally, our calculations have in some ways relied upon the idea that the critical exponents can be combined to form integer values. While this is true for all known examples of black holes, there are examples of systems for which this is not true. It may be interesting to explore whether our results can be extended to these cases as well.

\acknowledgements

RAH would like to thank Clifford Johnson for discussions on this work during a visit to USC in February 2020. The work of MCD and SLJ was supported by the University of Waterloo through the co-op programme. During the course of this work, RAH was supported in part by the Natural Sciences and Engineering Research Council of Canada  (NSERC), Asian Office of Aerospace Research and Development
Grant No. FA2386-19-1-4077, and received the support of a fellowship from ``la Caixa'' Foundation (ID 100010434) and from the European Union’s Horizon 2020 research and innovation
programme under the Marie Skłodowska-Curie grant agreement No 847648 under fellowship
code LCF/BQ/PI21/11830027. The work of RBM is supported by NSERC.

\appendix

\section{Example 3: Einstein-Gauss-Bonnet-Maxwell Gravity}

In this appendix, we test our results against charged black holes in Einstein-Gauss-Bonnet theory. These belong to Lovelock theory with $k_{\rm max} = 2$ and 
\be 
\mathcal{L}_m = - 4 \pi  F_{ab} F^{ab} \, ,
\ee
with $F_{ab}$ the Maxwell field strength. These black holes were considered in detail in~\cite{ Cai:2013qga,Frassino:2014pha}, where it was shown that the mass and equation of state have the form
\begin{align}
M=&\frac{\Sigma_{d-2}^{(\kappa)}(d-2)}{16\pi}\sum_{k=0}^{k=2}\alpha_k\kappa^kr_+^{d-1-2k}+\frac{\Sigma_{d-2}^{(\kappa)}Q^2}{2(d-3)r_+^{d-3}} \, ,
\\
P=&\frac{(d-2)T}{4 r_+}-\frac{(d-2)(d-3)}{16 \pi r_+^2}+\frac{(d-2)\alpha_2 T}{2 r_+^3} -\frac{(d-2)(d-5)\alpha_2}{16 \pi r_+^4}+\frac{Q^2}{2 r_+^{2(d-2)}} \, .
\end{align}

Our aim is to reproduce the calculations done for the RN-AdS case above for this class of black hole, for which the equation of state is considerably more complicated.  We will confirm that the expression derived earlier for the efficiency holds.  The equation of state has a critical point satisfying
\begin{align}
\frac{\partial P}{\partial r_+}=\frac{\partial ^2P}{\partial r_+^2}=0.
\end{align} 
We can solve the first partial derivative for the critical temperature $T_c$ to get
\be\label{critical_temperature_GB}
T_c=\frac{(d-3)r_c^2+{2(d-5)\alpha_2}-8\pi Q^2r_c^{8-2d}}{2\pi r_c(r_c^2+6\alpha_2)},
\ee
where $r_c$ is the critical horizon radius. However, the second partial derivative, which determines $r_c$, cannot be solved for analytically. In general dimensions, it takes the form
\begin{align}
&8\pi Q^2 r_c^8\left((5-2d)r_c^2+6(7-2d)\alpha_2\right)+r_c^{2d}\left((d-3)r_c^4-12r_c^2\alpha_2+12(d-5)\alpha_2^2\right)=0 \, ,
\end{align}
and we will impose this as a constraint in the calculations below. We can now express the critical pressure $P_c$ in terms of $T_c$ and the (as yet undetermined) $r_c$ by substituting our value for $T_c$ from equation \eqref{critical_temperature_GB} into the equation of state \eqref{Lovelock_EOS}. At this point, the expressions for general $d$ become too lengthy to present here. Therefore, we will present the discussion for the case $d=5$ and comment on the general case at the end. In $d=5$ we obtain
\begin{align}
T_c&=\frac{r_c^4-4\pi Q^2}{\pi r_c^3(r_c^2+6\alpha_2)}\, ,\\
P_c&=\frac{3 r_c^6-6 r_c^4 \alpha_2-4 \pi Q^2(5r_c^2+6\alpha_2)}{8\pi r_c^6(r_c^2+6\alpha_2)} 
\end{align}
for   the critical temperature and pressure respectively.

We are again considering the same set-up of the engine cycle, with the bottom right hand corner coincident with the critical point. The small $x$ expansion of $M_2-M_1$ is given in $d=5$ as
\begin{align}
M_2-M_1&=\frac{1}{16(V_c+3\sqrt{2V_c}\pi\alpha_2)} \bigg{[}3 \sqrt{2}V_c^{\frac{3}{2}}(y+2)-6\pi V_c(y-2)\alpha_2-2\sqrt{2V_c}\pi^3Q^2(5y+6)
\nonumber\\
&-12\pi^4Q^2(y+2)\alpha_2\bigg{]}x+\frac{3\sqrt{2}(V_c-2\pi^3Q^2)}{64\sqrt{V_c}}x^2 + \mathcal{O}(x^3)
\end{align}
where we have used the relation (\ref{volume_radius}) to replace $r_c$ in favour of $V_c$. The efficiency \eqref{eqn:efficiencyformula} can then be found similarly to the Lovelock example in the main text;  however even in $d=5$ the expression is too long to be included here.

One proceeds in a similar way as   in section \ref{sec:RN_AdS} by finding the Carnot efficiency $\eta_C$, and the ratio of $\eta/\eta_C$;   in the small $x$ limit we find 
\begin{align}\label{GB_eff_ratio}
\frac{\eta}{\eta_c}=&1-\frac{3\left(2\pi^3Q^2 - V_c\right)(\sqrt{2 V_c}+6\pi\alpha_2) x}{4\bigg[-3\sqrt{2}V_c^{\frac{3}{2}}(y+2)+6\pi V_c(y-2)\alpha_2+2\sqrt{2V_c}\pi^3Q^2(5y+6)+12\pi^4Q^2(y+2)\alpha_2\bigg]} 
\nn\\
&+\mathcal{O}(x^2).
\end{align}
We now continue our comparison with \eqref{eqn:normal_eff} by computing the relevant quantities $B_0$ and $A$. To do this we take the near-critical expansion of the equation of state which is given by
\be\label{near_crit_GB}
\rho=A\tau+B_0\tau\omega+f(V_c)\omega^2+C_0\omega^3,
\ee
with
\begin{align}
A=2-\frac{8(\sqrt{2V_c}\pi^3Q^2+3\pi V_c\alpha_2)}{-3\sqrt{2}V_c^{3/2}+6\pi V_c\alpha_2+10\sqrt{2V_c}\pi^3Q^2+12\pi^4Q^2\alpha_2},\\
B_0=-\frac{3(2\pi^3Q^2-V_c)(\sqrt{2V_c}+6\pi\alpha_2)}{2(-3\sqrt{2}V_c^{3/2}+6\pi V_c\alpha_2+10\sqrt{2V_c}\pi^3Q^2+12\pi^4Q^2\alpha_2)},\\
C_0=\frac{-15\sqrt{2}V_c^{3/2}+102\pi V_c\alpha_2+190\sqrt{2V_c}\pi^3Q^2+756\pi^4Q^2\alpha_2}{64\left(-3\sqrt{2}V_c^{3/2}+6\pi V_c\alpha_2+10\sqrt{2V_c}\pi^3Q^2+12\pi^4Q^2\alpha_2\right)}.
\end{align}

The as yet unimposed criticality condition $\partial ^2 P/\partial r_+^2=0$ is equivalent to $f(V_c)=0$ and so we can drop the $\omega^2$ term. Thus, as the critical point is governed by the standard critical exponents~\eqref{crit_exp_standard}, equation~\eqref{near_crit_GB} is identical to equation~\eqref{eqn:near_crit}.  As expected, the coefficient of $x$ in \eqref{GB_eff_ratio} can be manipulated such that it is equal to the $(-x)^{\delta - \frac{1}{\beta}}$ coefficient in~\eqref{eqn:ratio_gen}, confirming our results.

Although we have here only presented the argument of equality for the Gauss Bonnet case in $d=5$, we have verified in Maple that the argument works for Gauss-Bonnet gravity in all dimensions and similarly for third order Lovelock gravity in all dimensions.

\bibliography{mybib}
\end{document}